\documentclass[final,5p,times,twocolumn]{elsarticle}
\usepackage{graphicx}
\usepackage{amssymb}

\begin{document}

\begin{frontmatter}

\title{Transitions between levels of a quantum bouncer
induced by a noise-like perturbation}

\author[ILL]{C.~Codau}
\author[ILL]{V.V.~Nesvizhevsky}
\author[ILL,PSI]{M.~Fertl}
\author[LPSC]{G. Pignol}
\author[LPSC]{K.V. Protasov}

\address[ILL]{ILL, 6 rue Jules Horowitz, Grenoble, F-38042, France}
\address[LPSC]{Laboratoire de Physique Subatomique et de Cosmologie,
Universit\'e Joseph Fourier--CNRS/IN2P3--INPG, Grenoble, France}
\address[PSI]{Paul Scherrer Institut, 5232 Villigen, Switzerland}

\begin{abstract}
The probability of transition between levels of a quantum bouncer, induced by a noise-like perturbation, is calculated. 
The results are applied to two sources of noise (vibrations and mirror surface waviness) which might play an important role in future GRANIT experiment, 
aiming at precision studies of/with the neutron quantum bouncer. 
\end{abstract}

\end{frontmatter}

\section{Introduction}
\label{introduction}

The quantum problem of a ball bouncing above an ideal mirror was considered a long time ago but it was thought as a mere exercise in elementary quantum mechanics (see, for instance, \cite{Goldman}). 
Things did change since the quantization of energy levels of Ultra Cold Neutrons (UCN) bouncing above a mirror in the Earth's gravitational field has been demonstrated in an experiment performed at the Institute Laue Langevin (ILL) \cite{Nesvizhevsky:2002ef, Nesvizhevsky:2003ww, Nesvizhevsky:2005jc}.
Together with neutron interferometric measurements in the gravity field \cite{COW} and studies of bouncing cold atoms \cite{Dalibard,Peters}, bouncing neutrons provide one of the few situations where quantum effects in a gravity field can be observed. 

Measuring more accurately the parameters of the quantum levels would allow to search for new short-range interactions (spin-independent or spin-dependent), 
to study the interaction of a quantum system with a gravitational field, to check the equivalence principle in the quantum regime and to study various quantum optics phenomena. 
Recent reviews on the subject can be found in \cite{reviewBaessler,reviewNesvizhevski}. 
New experiments are being developed in this direction \cite{Sanuki,abele,Kreuz}, with the ultimate goal of performing the precise spectroscopy of the discrete energy levels \cite{NOVA,abele2}. 
In particular, we will focus on the second generation GRANIT spectrometer \cite{Kreuz} (GRAvitational Neutron Induced Transitions) that is being set up in Grenoble, with the aim of storing neutrons in quantum states for very long time. 

The principal element in the spectrometer is an horizontal mirror. 
In the given setup it is of square shape with a size of 30~cm and four vertical side walls constituting a trap for the gravitational quantum states. 
It should provide large lifetimes for the gravitationally bound quantum states of neutrons in the storage measuring mode. 
Extremely severe constraints for parameters of this mirror trap were identified in a detailed analysis \cite{Guillaume}. 
Some parameters have already been tested \cite{Nesvizhevsky2006:sap,Nesvizhevsky2007:m}, motivating the choice of the mirror's material. 

Although the quantum states of the bouncing neutrons are fundaamentally stable \cite{graviton},  various physical effects can induce parasitic transitions. 
Among these effects, vibrations and waviness of the mirror surface are of vital importance for the future experiment. 
These two effects can be taken into account within a common approach presented in this paper. 
An early analysis can be found in \cite{Dubna}.

Main sources of external vibrations in GRANIT come from large pumps of the ILL reactor. Important contribution could come, under certain conditions, from pumps of the spectrometer. Occasional large-amplitude low-frequency vibrations originate from large ILL cranes. Vibrations coming from numerous experimental installations and people in vicinity of GRANIT, also standard seismic noise, should not be ignored as well.               

\bigskip

First we will set notations for the quantum bouncer problem, focusing on its physical implementation, that is, neutrons falling in the Earth's gravitational field above a perfect mirror. The second section is devoted to a general expression for the probability of transition between two quantum levels, induced by a noise-like perturbation, which can be obtained within the first Born approximation (largely sufficient for our experimental needs). The third section treats a case of vibrations, for which first experimental measurements of mechanical noise on the future GRANIT position in the ILL reactor hall are presented. The fourth section does a case of a wavy mirror surface. Conclusions summarize the obtained results and give predictions for the future trap prepared for the GRANIT experiment.

\section{The quantum bouncer and transitions induced by small vibrations}
\label{part2}

Let us remind briefly the solution for the quantum bouncer problem \cite{Nesvizhevsky:2003ww}.
The wave function $\psi(z)$ of a quantum bouncer with the mass $m$ obeys the stationary Schr\"odinger equation
for the vertical motion along $z$ axis
\begin{equation}
-\frac{\hbar^2}{2 m} \frac{d^2 \psi}{dz^2} + m g z \ \psi = E \ \psi.
\label{schrodinger}
\end{equation}
The boundary condition $\psi(z = 0) = 0$ is due to the presence of a perfect mirror at height $z = 0$.
Let us now introduce the characteristic length and the characteristic energy of this problem, together with their value for the falling neutron case:
\begin{eqnarray}
z_0 & = \ \left( \frac{\hbar ^2}{2 m^2 g} \right)^{1/3} & = \ 5.87 \ \mu \rm{m} \\
E_0 & = \ \ m g z_0 & = \ 0.60 \ \rm{peV}.
\end{eqnarray}
The eigenproblem (\ref{schrodinger}) can be solved in terms of the first Airy function ${\rm Ai}(X)$, which have an infinite number of negative zeros, denoted by $-\lambda_1, -\lambda_2, \dots$ in decreasing order. 
The stationary states energies are $E_n = E_0 \lambda_n$, and the wave function of the $n^{th}$ (non-degenerate) state reads
\begin{equation}
\psi_n(z) = C_n \mbox{Ai~} \left(\frac{z-z_n}{z_0} \right) \ \theta(z)
\end{equation}
where $z_n = z_0 \lambda_n$, and $C_n$ normalizes the probability to find the neutron anywhere to $1$.
The sequence of the zeros of the Airy function has no simple analytic expression, but while applying the Bohr-Sommerfeld rules giving approximate energy levels, we find a fairly good approximation for this sequence:
\begin{equation}
\lambda_n \approx \left( \frac{3 \pi}{8}(4n-1) \right)^{2/3}. 
\end{equation}
It is accurate within 1 \%, even for $n=1$, and is exact in the semi-classical limit ($n \rightarrow \infty$).
Table \ref{frequencies} gives the expected transition frequencies $2 \pi f_{nm} = \left( E_n - E_m \right)/\hbar$ following exact calculation of the $\lambda_n$ sequence. 

\begin{table}
\begin{center}
\begin{tabular}{c|llllll}
$f_{nm}$ [Hz]	& 1     & 2     & 3     & 4     & 5     & 6	\\
\hline
1		& 0	& 254	& 462	& 645	& 813	& 969	\\
2		&	& 0	& 208	& 391	& 559	& 716	\\
3		&	&	& 0	& 184	& 351	& 508	\\
4		&	&	&	& 0	& 168	& 324	\\
5		&	&	&	&	& 0	& 156
\end{tabular}
\end{center}
\caption{\label{frequencies} Expected transition frequencies for the six lowest gravitational neutron quantum states ; $g = 9.806$~m/s$^2$.}
\end{table}

In this article we are concerned with the problem of stability of the neutron quantum states against vibrations of the bottom horizontal mirror. 
This problem can also be seen as horizontal motion of a neutron above a rough mirror. 
From this point of view these two effects are of the same nature and can be treated within a common approach.


Let us suppose that the mirror vibrates with a time dependent height described by a function $\zeta (t)$. 
Following the approach in \cite{refMey}, we perform a transformation
\begin{eqnarray}
Z = z - \zeta (t) \quad \phi(Z, t) = \psi(z,t)
\end{eqnarray}
(other coordinates are unchanged). 
The new wavefunction satisfies a modified Schr\"odinger equation
\begin{eqnarray}
\nonumber
i \hbar \frac{\partial \phi}{\partial t}(Z, t) & = & \left(-\frac{\hbar^2}{2m} \frac{\partial^2}{\partial Z^2} + mgZ \right) \phi(Z,t) \\ 
					&	 + & \left( mg\zeta(t) + i \hbar \zeta'(t) \frac{\partial}{\partial Z} \right) \phi(Z, t) \\
\nonumber
\phi(0, t) & = & 0
\end{eqnarray} 
This transformation thus introduces an effective perturbation potential
\begin{eqnarray}
\widehat V = mg\zeta (t) - \zeta'(t) \widehat p_z.
\end{eqnarray}

Let us note that the first term, being $z$-independent, 
provides no transition between quantum levels (all levels are shifted simultaneously).
The relevant time-dependent perturbation term $- \zeta'(t) \widehat p_z$ 
corresponds to the transfer of horizontal motion
into vertical one during a collision with the moving mirror. 
The (spatial) matrix element calculated with two wave functions $\Psi
_k^{(0)}  = \psi _k (z)e^{ - 2 i \pi f_k t} $ is
equal to
\begin{eqnarray}
\nonumber
V_{nm} (t) &=& \int {\Psi _n^{*(0)} \widehat{V}(t)\Psi _m^{(0)} dz} \\
\nonumber
	& = &  - \zeta'(t) e^{2 i \pi f _{nm} t} \int {\psi _n^* \widehat p_z \psi
_m dz}  \\
&=&  - \zeta'(t) e^{2 i \pi f_{nm} t} \left( {p_z } \right)_{nm}.
\end{eqnarray}

At first order in perturbation theory, the probability of the corresponding transition after an observation time $T$ is equal to
\begin{eqnarray}
\nonumber
P_{n \to m} (T) & = & \frac{1}{{\hbar ^2 }}\left|
{\int\limits_{0}^{T} {V_{nm} (t)dt} } \right|^2  \\ 
 & = &
\frac{{\left( {p_z } \right)_{nm}^2 }}{{\hbar ^2 }}\left|
{\int\limits_{0}^{T} {\zeta'(t) e^{2 i \pi f _{nm} t} dt} } \right|^2 .
\end{eqnarray}

The matrix element $\left( {p_z }\right)_{nm}$ can be easily calculated (see Appendix)
and one thus obtains the probability of transition between two states per unit of time
in the form
\begin{equation}
\label{transitionProb}
p_{n \to m}^{\mbox{\tiny sys}}  = \left(\frac{mg}{2\pi f_{nm} \hbar}\right)^2 \ \lim_{T \rightarrow \infty} \frac{1}{T} \left| \int_{0}^{T} {\zeta'(t) e^{2 i \pi f _{nm} t} dt} \right|^2.
\label{probtrans}
\end{equation}
This general formula makes apparent that nonresonant transitions between quantum states are induced 
by the spectral component of the noise at the frequency corresponding to transitions from an initial quantum state to a closest level.

\section{Experimentally measured spectrum of vibrations at the installation}
\label{vibrations}

\begin{figure}
\begin{center}
\includegraphics[width=0.95\linewidth]{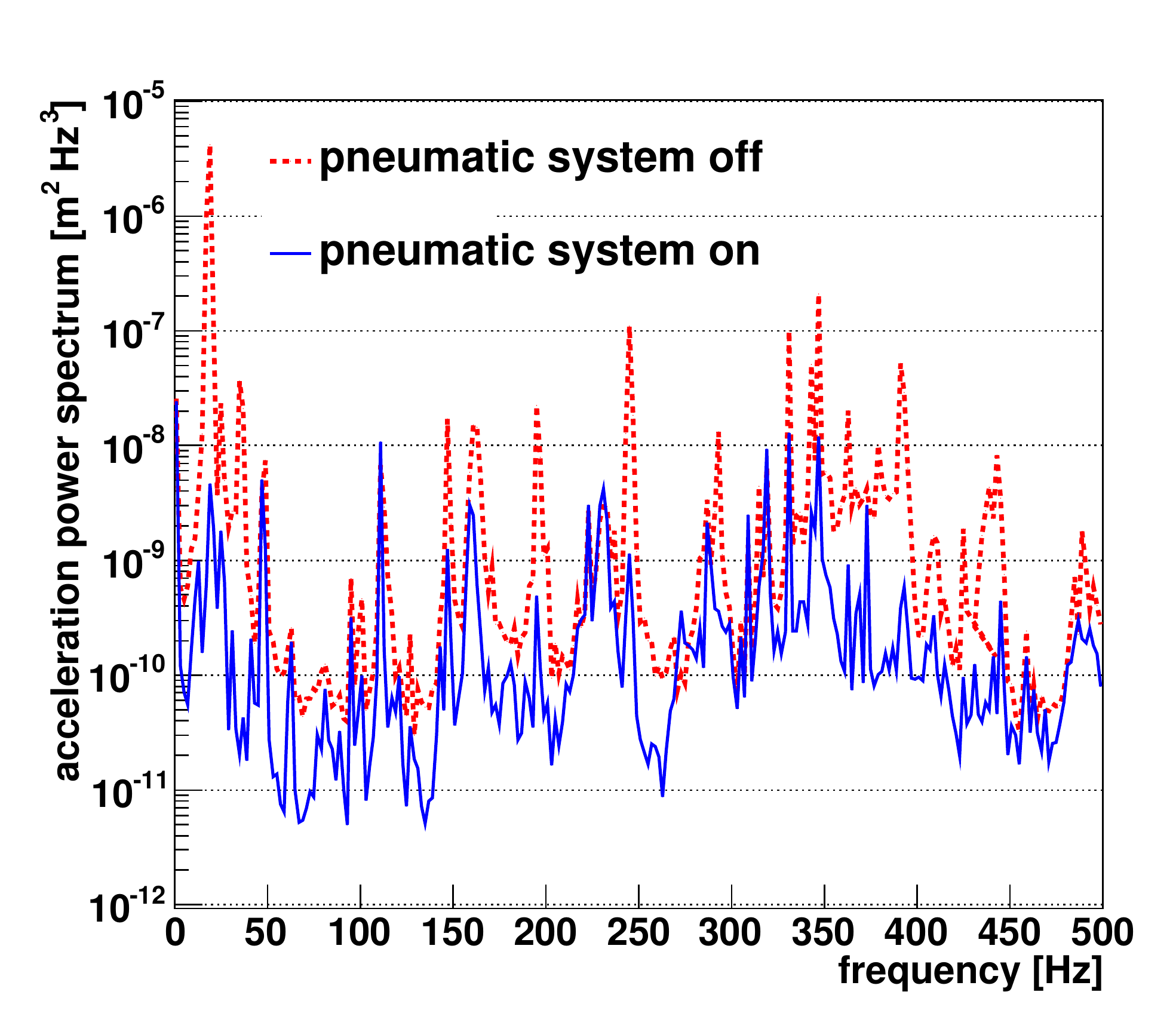}
\caption{Power spectra of vibrations measured inside the GRANIT spectrometer in realistic conditions, with and without the passive pneumatic compensation.} 
\label{vibrationPowerSpectrum}
\end{center}
\end{figure}

A series of measurement have been performed to characterize the level of vibration noise in the GRANIT spectrometer. 
We used an accelerometer sensitive in the frequency range from 0 to 500~Hz; precisely the range of interest in the first experiments according to table \ref{frequencies}. 
A low frequency accelerometer, sensitive in the range from 0 to 100 Hz, was also used for consistency checks. 
The calibration values given by the manufacturer were further checked successfully in measuring the accelerometers signals while doing controlled low frequency harmonic motion of the devices. 
It is important to mention that the ILL reactor, as well as the spectrometer vacuum pumps, were on during the measurements. 
Thus, the measured vibration level is close to that in realistic experimental conditions, apart from a possible contribution of the UCN source. 
This specific contribution will be studied when possible. 
The accelerometers are placed inside the GRANIT apparatus and sample the vertical acceleration $a(t) = \zeta''(t)$ from which we can calculate the \emph{acceleration power spectrum}
\begin{equation}
S_a(f) = \lim_{T \rightarrow \infty} \frac{1}{T} \left| \int_{0}^{T} {a(t) e^{2 i \pi f t} dt} \right|^2.
\end{equation}
The acceleration power spectrum is related to the displacement acceleration spectrum by $S_a(f) = (2\pi f)^4 S(f)$.
Figure \ref{vibrationPowerSpectrum} shows the measured acceleration power spectrum; the situation with and without the pneumatic feet system are compared. 
The pneumatic system suppresses the vibration power by about one order of magnitude, thus increasing the quantum levels lifetime by the same factor. 

Now from these measurements and the general formalism described in the previous section we can calculate the expected lifetimes of the quantum levels. 
In terms of the vertical acceleration of the mirror the general formula (\ref{transitionProb}) becomes 
\begin{equation}
\label{ProbVibrations}
p_{n \to m}^{\rm vibrations}  = \left(\frac{mg}{\hbar}\right)^2 \ \frac{1}{ ( 2\pi f_{nm})^4 } S_a(f_{nm}).
\end{equation}
The quantum levels lifetimes are calculated, summing the contributions (\ref{ProbVibrations}) for all transitions. Namely, for the $n^{\rm th}$ level:
\begin{equation}
\label{LifetimeVibrations}
\frac{1}{T_n^{\rm vibrations}} = \sum_{m \neq n} p_{n \rightarrow m}^{\rm vibrations}
\end{equation}
Figure \ref{lifetimes} shows the expected lifetime of the 50 lowest quantum states. They are longer than the $\beta$ decay time of the neutron of about 900~s. 
Effects of crane motion and human activity around the experiment have also been investigated, and are found not to be a concern for the stability of the neutron gravitational quantum states. 
Only deliberate shocks against the apparatus would spoil the experiment.

\section{Surface waviness}
\label{waviness}

An analogous approach can be used for the surface noise induced by waviness of the mirror.
In this case, the mirror surface as a function of horizontal coordinate $x$ is described
by the function $\zeta (x)$. 
In its own frame, a neutron with the horizontal velocity $v$ (of about 5~m/s) 
sees the perturbation potential $- v \zeta'(v t) \widehat p_z$.

For the probability of transition between two states per unit of time
one obtains the same expression as that in (\ref{probtrans})
\begin{eqnarray}
p_{n \to m}^{\mbox{\tiny surf}}  = \frac{1}{v} \left(\frac{mg}{\hbar ^2 }\right)^2 S\left(\frac{f_{nm}}{v}\right),
\end{eqnarray}
 where $S(K)$ denotes the spectral density of surface noise
\begin{eqnarray}
S(K) = \mathop {\lim }\limits_{L \to \infty }
\frac{1}{L}\left| {\int\limits_{0}^{L} {\zeta (x)e^{i2\pi Kx} dx} } \right|^2.
\end{eqnarray}

Note that for lowest quantum states, the characteristic spacial scale is 
250 Hz/(5~m/s)= 50~m$^{-1}$.
So it is more appropriate to speak in terms of waviness than surface roughness.

\begin{figure}
\begin{center}
\includegraphics[width=0.95\linewidth]{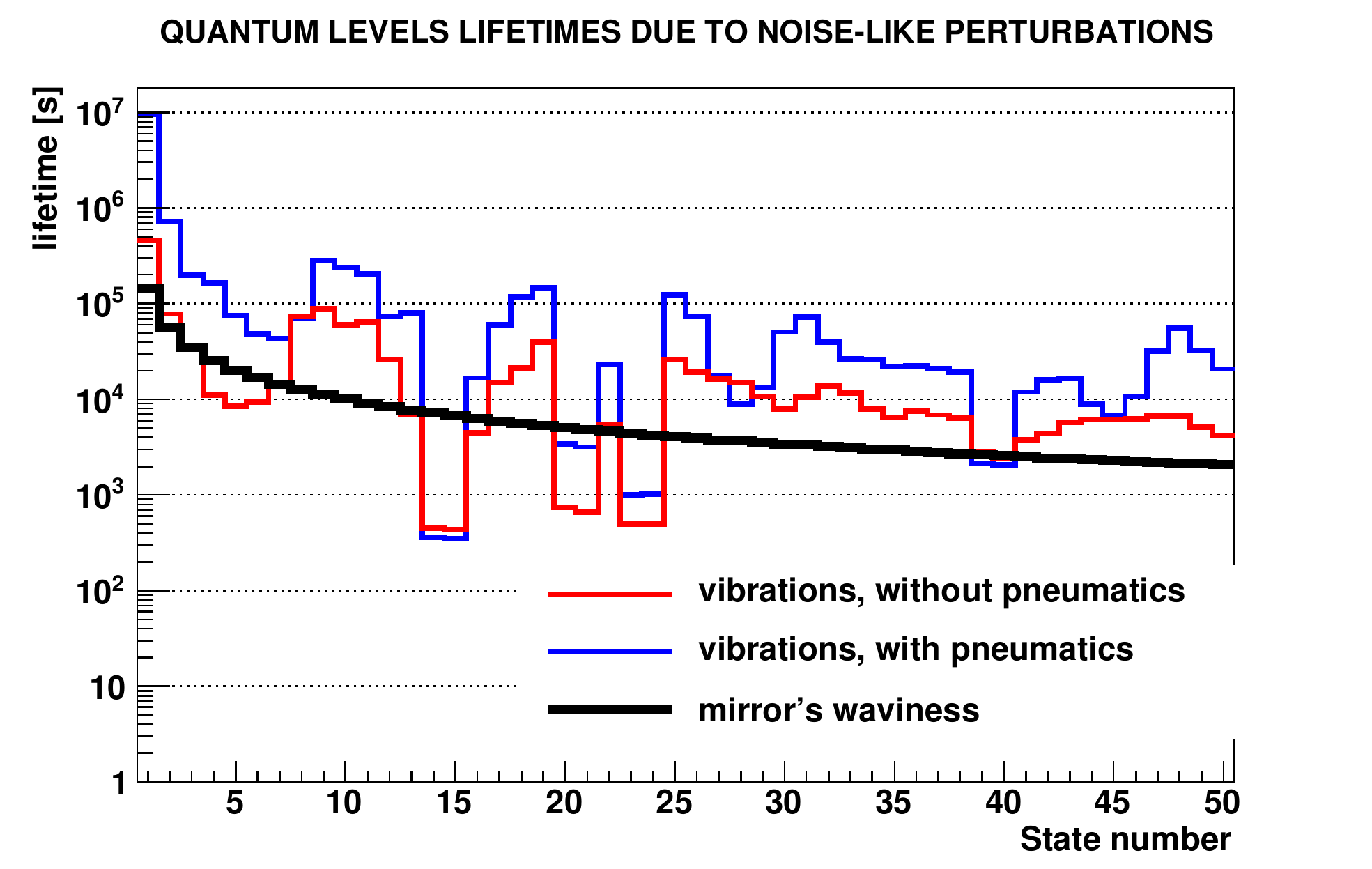}
\caption{Expected lifetimes of neutron gravitational quantum states due to vibrations of the mirrors and waviness of the mirror. 
The vibration contribution is estimated from measured in-situ vibration level with and without the passive pneumatic compensation. } 
\label{lifetimes}
\end{center}
\end{figure}

The spectral function $S(K)$ has not yet been measured for the future GRANIT mirrors. 
Nevertheless, it is interesting to evaluate the lifetimes of the neutrons in quantum states by using
spectral function known for other mirrors with properties close to that for GRANIT mirror. 
For instance, the spectral density function for silicon substrate (plates of 300 mm diameter) was measured using Long Trace Profiler in \cite{Siquality}:
\begin{eqnarray}
\label{PSDwaviness}
S(K) = 2 \times 10^{-4} \left(\frac{K}{1 \mbox{ mm}^{-1}}\right)^{-2.9}\mbox{nm$^2$ mm}, 
\end{eqnarray}
in the range $5 \, {\rm m}^{-1} < K < 500 \, {\rm m}^{-1}$, thus including the range around $K \approx 50 \, {\rm m}^{-1}$ important for GRANIT. 
In turn one can calculate the expected lifetimes assuming the GRANIT mirror will be as good as that in (\ref{PSDwaviness}) by summing the contribution for all transitions as already done for the vibration case in eq. (\ref{LifetimeVibrations}). 
The result is shown in fig. \ref{lifetimes}. 
Again the expected lifetimes are longer than the neutron $\beta$ decay lifetime. 

\section{Conclusion}
\label{conclusion}

We have developed a general theoretical formalism describing transitions between levels of a quantum bouncer induced by noise-like perturbations. 
It is applied to transitions caused by vibrations and mirror waviness in the GRANIT experiment. 

The GRANIT mirror trap is in production by SESO at the moment. 
Therefore its spectral roughness (waviness) density was estimated using published data on an analogous mirror. 
We are convinced that mirror waviness will not produce major false transitions.  

Vibration spectra were measured in various GRANIT configurations. We showed that vibration, when properly controlled on-line, would not disturb the first measurement with GRANIT. If vibrations exceed temporarily some defined threshold, the measurement will be interrupted. 
One should note that lifetimes of low states are longer by orders of magnitude than those of excited states. This is explained by higher resonant frequencies for lower states; both the probability of transition and the spectral noise density are low at high frequency. In order to profit from this argument we increased the total weight of our spectrometer by an order of magnitude compared with its previous version; another large improvement consists of much lower vibrations at new position of GRANIT at the level C in the ILL reactor.   

\section*{Acknowledgments}
The authors are grateful to members of the GRANIT collaboration for help and advice.

\section*{Appendix}

To calculate the matrix element $\left( {p_z }\right)_{nm}$, one can multiply the 
Schr\"odinger equation (\ref{schrodinger}) for the wave function $\psi_m$
\begin{equation}
-\frac{\hbar^2}{2 m} \frac{d^2 \psi_m}{dz^2} + m g z \ \psi_m = E_m \ \psi_m
\end{equation}
by $\psi_n'$, multiply the Schr\"odinger equation for the wave function $\psi_n$ by $\psi_m'$,
to sum them and to calculate the integral over $z$ variable between 0 and $\infty$.
One obtains,
\begin{eqnarray*}
&&\int_0^\infty\left(\psi_n''\psi_m'+\psi_m''\psi_n'\right)dz+\frac{2m}{\hbar^2}
\int_0^\infty\left(E_n\psi_n\psi_m'+E_m\psi_m\psi_n'\right)dz\\
&&=\frac{2m^2g}{\hbar^2}\int_0^\infty z\left(\psi_n\psi_m'+\psi_m\psi_n'\right)dz.
\end{eqnarray*}
The first integral is equal to
\begin{eqnarray*}
\int_0^\infty\left(\psi_n''\psi_m'+\psi_m''\psi_n'\right)dz
=\psi_n'(0)\psi_m'(0)=\frac{1}{z_0^3}=\frac{2m^2g}{\hbar^2}.
\end{eqnarray*}
The second one
\begin{eqnarray*}
\int_0^\infty\left(E_n\psi_n\psi_m'+E_m\psi_m\psi_n'\right)dz &=& (E_n-E_m)
\int_0^\infty\psi_n\psi_m'dz \\
&=&2\pi i f_{mn} \left( {p_z }\right)_{nm}.
\end{eqnarray*}
The third one is equal to
\begin{eqnarray*}
\int_0^\infty z\left(\psi_n\psi_m'+\psi_m\psi_n'\right)dz = \int_0^\infty \frac{d}{dz}\left(z\psi_n\psi_m\right)dz=0.
\end{eqnarray*}

Thus, for the matrix element $\left( {p_z }\right)_{nm}$, one obtains
\begin{eqnarray*}
\left( {p_z }\right)_{nm}=i\frac{mg}{2\pi}\frac{1}{f_{mn}}.
\end{eqnarray*}

\end{document}